\documentclass[a4paper, 10pt, oneside]{article}

\usepackage[utf8x]{inputenc}
\usepackage[T1]{fontenc}
\usepackage[UKenglish]{babel}
\usepackage{lmodern}
\usepackage{IEEEtrantools}
\usepackage{mathtools}
\usepackage{amsmath}
\usepackage{amssymb}
\usepackage{amsthm}
\usepackage{algorithm}
\usepackage{algpseudocode}
\usepackage{tikz}
\usepackage{comment}
\usepackage[bottom]{footmisc}
\usepackage[margin=2.5cm]{geometry}
\usepackage{fancyhdr}
\usepackage{hyperref}

\theoremstyle{plain}
\newtheorem{theorem}{Theorem}
\newtheorem{lemma}[theorem]{Lemma}
\newtheorem{corollary}[theorem]{Corollary}

\theoremstyle{definition}
\newtheorem{definition}[theorem]{Definition}

\newtheorem{remark}[theorem]{Remark}

\numberwithin{equation}{section}

\newcommand{\bbR}{\mathbb{R}}
\newcommand{\bbN}{\mathbb{N}}
\newcommand{\bbZ}{\mathbb{Z}}

\newcommand{\calO}{\mathcal{O}}
\newcommand{\calC}{\mathcal{C}}
\newcommand{\calR}{\mathcal{R}}
\newcommand{\equalDef}{\coloneqq}
\newcommand{\assign}{\coloneqq}

\newcommand{\compLeq}[1]{\calO\left( #1 \right)}
\newcommand{\compGeq}[1]{\Omega\left( #1 \right)}

\newcommand{\copt}{C_{\mathrm{opt}}}
\newcommand{\popt}{P_{\mathrm{opt}}}

\DeclareMathOperator*{\argmin}{argmin}

\providecommand{\ceilGraph}[1]{G_{#1}^{+}}
\providecommand{\floorGraph}[1]{G_{#1}^{-}}
\providecommand{\ceilCost}[1]{c_{#1}^{+}}
\providecommand{\floorCost}[1]{c_{#1}^{-}}
\providecommand{\rowtime}[1]{f_{#1}}
\newcommand{\natOne}{\bbN^+}
\newcommand{\natZero}{\bbN_0}
\newcommand{\nonNegReals}{\bbR_0^+}
\newcommand{\integers}{\bbZ}
\newcommand{\posReals}{\bbR^+}

\begin{document}


\renewcommand*{\thefootnote}{\fnsymbol{footnote}}

\title{Improved Approximation Schemes for the Restricted Shortest Path Problem\footnotemark[1]}
\footnotetext[1]{The results presented in this paper (except for the improvement on planar graphs) were developed as a bachelor thesis \cite{holzmuller2016}.} 
\author{David Holzmüller\footnotemark[2]}
\footnotetext[2]{Corresponding author \\ Institut fuer Formale Methoden der Informatik (FMI), Universitaet Stuttgart, Universitätsstraße 38, 70619 Stuttgart, Baden-Wuerttemberg, Germany \\
  \texttt{david.holzmueller@online.de}}
  
\renewcommand*{\thefootnote}{\arabic{footnote}}
  



\maketitle 

\begin{abstract}

The Restricted Shortest Path (RSP) problem, also known as the Delay-Constrained Least-Cost (DCLC) problem, is an NP-hard bicriteria optimization problem on graphs with $n$ vertices and $m$ edges. In a graph where each edge is assigned a cost and a delay, the goal is to find a min-cost path which does not exceed a delay bound. In this paper, we present improved approximation schemes for RSP on several graph classes. For planar graphs, undirected graphs with positive integer resource (= delay) values, and graphs with $m \in \compGeq{n \log n}$, we obtain $(1 + \varepsilon)$-approximations in time $\compLeq{mn/\varepsilon}$. For general graphs and directed acyclic graphs, we match the results by Xue et al. (2008, \cite{xue2008}) and Ergun et al. (2002, \cite{ergun2002}), respectively, but with arguably simpler algorithms.

\end{abstract}


\section{Introduction}

The Restricted Shortest Path (RSP) problem is concerned with finding shortest paths under a constraint, which is given by a length bound on a second edge weight function. While being of theoretical interest, the RSP problem also occurs in Quality of Service (QoS) routing problems in high-speed networks \cite{ergun2002}. Since the RSP problem is NP-hard \cite{garey1979}, efficient approximation algorithms are particularly interesting. A formal definition of the problem can be given as follows:

\begin{definition}
Let $G = (V, E, c, r)$ be a directed graph with two weight functions $c, r: E \rightarrow \nonNegReals$. Let $n \equalDef |V|$ and $m \equalDef |E|$. Without loss of generality, we assume $G$ to be weakly connected, hence $m \geq n-1$. For $e \in E$, we refer to $c(e)$ and $r(e)$ as the cost or resource consumption (e.g. delay) of an edge, respectively. Besides this graph, the input for the RSP problem consists of start and target nodes $s, t \in V$ ($s \neq t$) and a bound $R \in \nonNegReals$ imposed on the resource consumption of a path. For simplicity, we encode paths as sets of edges. Let $P_v$ be the set of all paths from $s$ to $v \in V$. We define the cost and the resource consumption of a path $p \in P_v$ by
\begin{IEEEeqnarray*}{+rCl+x*}
C_G(p) & \equalDef & \sum_{e \in p} c(e) \\
R_G(p) & \equalDef & \sum_{e \in p} r(e)~.
\end{IEEEeqnarray*}
While any path $p \in P_t$ satisfying $R_G(p) \leq R$ will be called a solution to the problem, we are interested in a (nearly) optimal solution. The optimal cost and the optimal solution are
\begin{IEEEeqnarray*}{*l+rCl+x*}
&\copt(G) & \equalDef & \min_{\substack{p \in P_t \\ R_G(p) \leq R}} C_G(p) \\
\text{and} \\
&\popt(G) & \equalDef & \argmin_{\substack{p \in P_t \\ R_G(p) \leq R}} C_G(p)~,
\end{IEEEeqnarray*}
where the latter might not be uniquely determined (but its cost is).
\end{definition}

In related papers, the resource consumption $r$ is often called delay. Like Xue et al. \cite{xue2008}, we assume a computational model where arithmetic operations run in time $\compLeq{1}$. 

\subsection{Related Work}
Hassin \cite{hassin1992} used a simple pseudopolynomial dynamic programming algorithm to solve the restricted shortest path problem exactly on graphs with $c(e) \in \natOne$ for all $e \in E$. Applying this algorithm to the original graph with scaled and rounded cost values, he obtained a $(1 + \varepsilon)$-approximation of the optimal solution in polynomial time. A $(1 + \varepsilon)$-approximation is a solution $p_\varepsilon$ satisfying $C_G(p_\varepsilon) \leq (1 + \varepsilon)\copt(G)$. Lorenz and Raz \cite{lorenz2001} improved Hassin's result, presenting an approximation scheme with a runtime complexity of $\compLeq{nm(1/\varepsilon + \log \log n)}$. Ergun et al. \cite{ergun2002} showed that a variant of this algorithm can be applied to directed acyclic graphs, yielding a runtime complexity of $\compLeq{nm/\varepsilon}$.\footnote{In the following, we assume $\varepsilon < C$ for some constant $C$, which means that $\compLeq{nm/\varepsilon}$ is equivalent to the more precise $\compLeq{nm(1/\varepsilon + 1)}$.} Xue et al. \cite{xue2008} used an extended dynamic programming scheme to handle edges with cost zero and introduced a second binary search phase in the algorithm by Lorenz and Raz \cite{lorenz2001} to obtain an algorithm with a time complexity of $\compLeq{nm(1/\varepsilon + \log \log \log n)}$. 

(Fully) Polynomial Time Approximation Schemes based on dynamic
programming with scaling often follow a two-step pattern: first obtain a
coarse approximation for the problem, and then use this coarse
approximation to determine a suitable scaling factor and solve the
scaled problem via dynamic programming. Typically, the second step
(solution of the scaled problem via DP) dominates the overall running
time. For example, in the classical FPTAS for the knapsack problem a
constant approximation is determined in $\compLeq{n \log n}$ time, based on which
the scaled instance is solved via DP in $\compLeq{n^2/\varepsilon}$ time.

There is a slight oddity for the RSP problem that for the known FPTA
schemes  the determination of the right scaling factor in fact dominates
the running time (for not too small values of $\varepsilon$) rather than the
supposedly more difficult solution of the DP ($\compLeq{nm \log \log \log n}$ vs
$\compLeq{nm/\varepsilon}$). Hence our result can also be interpreted as removing this
oddity for some graph classes.

\subsection{Contribution}
In this paper, we modify the algorithm by Xue et al. \cite{xue2008}, replacing the binary search they used by a linear search. This does not only achieve the same runtime complexity but it also allows adaptions for more efficient algorithms on several types of graphs.\footnote{While the adaptions might have also been possible using binary search, we consider the linear search variant to be simpler.} The complexity results of the algorithms in this paper are summarized in Table \ref{table:complexity} and compared to the best previous results we are aware of.

\begin{table}[hbt]
\begin{center}
\caption{Complexity results presented in this paper and previous results} \label{table:complexity}
\begin{tabular}{|c|c|c|}
\hline
Graphs & Best previous result & Our result \\
\hline
all & $\compLeq{nm(1/\varepsilon + \log \log \log n)}$ \cite{xue2008} & \begin{tabular}{c}
$\compLeq{nm(1/\varepsilon + \frac{n \log n}{m})}$ \\
$\compLeq{nm(1/\varepsilon + \log \log \log n)}$
\end{tabular}
\\
\hline
acyclic & $\compLeq{nm/\varepsilon}$ \cite{ergun2002} & $\compLeq{nm/\varepsilon}$ \\
\hline
planar & & $\compLeq{nm/\varepsilon}$ \\
\hline
undirected, $r(e) \in \natOne$ & & $\compLeq{nm/\varepsilon}$ \\
\hline
$r(e) \in \natZero$ & & $\compLeq{nm(1/\varepsilon + \frac{n}{m}\log \log R)}$ \\
\hline
\end{tabular}
\end{center}
\end{table}

Since our algorithm uses parts of the algorithms by Xue et al. \cite{xue2008} and Lorenz and Raz \cite{lorenz2001}, we will present the relevant parts here in a reformulation that suits our purpose. In the following, we explain the algorithms used by Xue et al. to find an optimal solution for integer-cost RSP, adding some remarks about special graph classes. Then, we will outline the algorithm by Lorenz and Raz \cite{lorenz2001} before presenting our results in section \ref{subsection:mainResults}. 

\section{Optimal Solution for Integer Cost} \label{section:optimalSolution}

Assuming $c(e) \in \natZero$ for all $e \in E$, we can solve the problem exactly using dynamic programming. We will first explain a simple algorithm for the case $c(e) \in \natOne$ proposed by Hassin \cite{hassin1992}. Afterwards, we will derive an extension for $c(e) \in \natZero$ which was (slightly differently) used by Xue et al. \cite{xue2008}. Let
\begin{IEEEeqnarray*}{+rCl+x*}
r_{v, i} & \equalDef & \min_{\substack{p \in P_v \\ C_G(p) \leq i}} R_G(p)
\end{IEEEeqnarray*}
for all $v \in V, i \in \integers$. Since the paths cannot have negative cost, we know that $r_{v, i} = \infty$ for $i < 0$ and $r_{s, i} = 0$ for $i \geq 0$. The other values obey the equation
\begin{IEEEeqnarray*}{+rCl+x*}
r_{v, i} & = & \min_{e = (u, v) \in E} r_{u, i - c(e)} + r(e)~.
\end{IEEEeqnarray*}
If $c(e) \geq 1$ for all $e \in E$, this directly translates to a dynamic programming scheme where a solution with cost $i$ exists iff $r_{t, i} \leq R$. If this condition is satisfied, the algorithm stops and traces the optimal path back from the computed table. At this time, we have tried $1 + \copt(G)$ values for $i$. For each of those, we had to compute $n$ values by considering $m \geq n-1$ edges in total. Thus, the time complexity is $\compLeq{m(1 + \copt(G))}$ for this case.

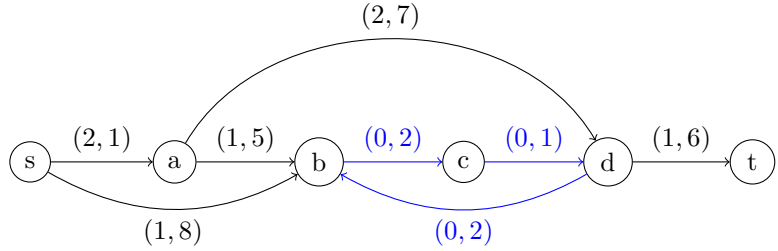
\begin{figure}[phbt]
\begin{center}
\begin{tikzpicture}[node distance = 1.9cm]
\node[circle, draw] (s) {s};
\node[circle, draw, right of = s] (a) {a};
\node[circle, draw, right of = a] (b) {b};
\node[circle, draw, right of = b] (c) {c};
\node[circle, draw, right of = c] (d) {d};
\node[circle, draw, right of = d] (t) {t};

\draw[->] (s) -- node[anchor=south] {$(2, 1)$} (a);
\draw[->] (a) -- node[anchor=south] {$(1, 5)$} (b);
\draw[->] (s) to [out=-30, in=-150] node[anchor=north] {$(1, 8)$} (b);
\draw[->] (a) to [out=60, in=120] node[anchor=south] {$(2, 7)$} (d);
\draw[->, color=blue] (b) -- node[anchor=south] {$(0, 2)$} (c);
\draw[->, color=blue] (c) -- node[anchor=south] {$(0, 1)$} (d);
\draw[->, color=blue] (d) to [out=-150, in=-30] node[anchor=north] {$(0, 2)$} (b);
\draw[->] (d) -- node[anchor=south] {$(1, 6)$} (t);
\end{tikzpicture}
\caption{Example graph $G_0$ including zero-cost edges. Each edge $e$ is labelled with $(c(e), r(e))$.} \label{fig:nnEx}
\end{center}
\end{figure}

\begin{figure}[hbt]
\begin{center}
\begin{tikzpicture}[node distance = 2.2cm]
\node[circle, draw] (sa) {$r_{s, 0}$};
\node[circle, draw, right of = sa] (aa) {$r_{a, 0}$};
\node[circle, draw, right of = aa] (ba) {$r_{b, 0}$};
\node[circle, draw, right of = ba] (ca) {$r_{c, 0}$};
\node[circle, draw, right of = ca] (da) {$r_{d, 0}$};
\node[circle, draw, right of = da] (ta) {$r_{t, 0}$};
\node[circle, draw, below of = sa] (sb) {$r_{s, 1}$};
\node[circle, draw, right of = sb] (ab) {$r_{a, 1}$};
\node[circle, draw, right of = ab] (bb) {$r_{b, 1}$};
\node[circle, draw, right of = bb] (cb) {$r_{c, 1}$};
\node[circle, draw, right of = cb] (db) {$r_{d, 1}$};
\node[circle, draw, right of = db] (tb) {$r_{t, 1}$};
\node[circle, draw, below of = sb] (sc) {$r_{s, 2}$};
\node[circle, draw, right of = sc] (ac) {$r_{a, 2}$};
\node[circle, draw, right of = ac] (bc) {$r_{b, 2}$};
\node[circle, draw, right of = bc] (cc) {$r_{c, 2}$};
\node[circle, draw, right of = cc] (dc) {$r_{d, 2}$};
\node[circle, draw, right of = dc] (tc) {$r_{t, 2}$};
\node[circle, draw, below of = sc] (sd) {$r_{s, 3}$};
\node[circle, draw, right of = sd] (ad) {$r_{a, 3}$};
\node[circle, draw, right of = ad] (bd) {$r_{b, 3}$};
\node[circle, draw, right of = bd] (cd) {$r_{c, 3}$};
\node[circle, draw, right of = cd] (dd) {$r_{d, 3}$};
\node[circle, draw, right of = dd] (td) {$r_{t, 3}$};
\node[circle, draw, below of = sd] (se) {$r_{s, 4}$};
\node[circle, draw, right of = se] (ae) {$r_{a, 4}$};
\node[circle, draw, right of = ae] (be) {$r_{b, 4}$};
\node[circle, draw, right of = be] (ce) {$r_{c, 4}$};
\node[circle, draw, right of = ce] (de) {$r_{d, 4}$};
\node[circle, draw, right of = de] (te) {$r_{t, 4}$};

\draw[->] (sa) -- node[anchor=west] {$1$} (ac);
\draw[->] (sb) -- node[anchor=west] {$1$} (ad);
\draw[->] (sc) -- node[anchor=west] {$1$} (ae);

\draw[->] (sa) -- node[anchor=south] {$8$} (bb);
\draw[->] (sb) -- node[anchor=south] {$8$} (bc);
\draw[->] (sc) -- node[anchor=south] {$8$} (bd);
\draw[->] (sd) -- node[anchor=south] {$8$} (be);

\draw[->] (aa) -- node[anchor=east] {$5$} (bb);
\draw[->] (ab) -- node[anchor=east] {$5$} (bc);
\draw[->] (ac) -- node[anchor=east] {$5$} (bd);
\draw[->] (ad) -- node[anchor=east] {$5$} (be);

\draw[->] (aa) -- node[anchor=south, xshift=-1.2cm, yshift=0.8cm] {$7$} (dc);
\draw[->] (ab) -- node[anchor=south, xshift=-1.2cm, yshift=0.8cm] {$7$} (dd);
\draw[->] (ac) -- node[anchor=south, xshift=-1.2cm, yshift=0.8cm] {$7$} (de);

\draw[->, color=blue] (ba) -- node[anchor=south, xshift=0.2cm] {$2$} (ca);
\draw[->, color=blue] (bb) -- node[anchor=south, xshift=0.2cm] {$2$} (cb);
\draw[->, color=blue] (bc) -- node[anchor=south, xshift=0.2cm] {$2$} (cc);
\draw[->, color=blue] (bd) -- node[anchor=south, xshift=0.2cm] {$2$} (cd);
\draw[->, color=blue] (be) -- node[anchor=south, xshift=0.2cm] {$2$} (ce);

\draw[->, color=blue] (ca) -- node[anchor=south] {$1$} (da);
\draw[->, color=blue] (cb) -- node[anchor=south] {$1$} (db);
\draw[->, color=blue] (cc) -- node[anchor=south] {$1$} (dc);
\draw[->, color=blue] (cd) -- node[anchor=south] {$1$} (dd);
\draw[->, color=blue] (ce) -- node[anchor=south] {$1$} (de);

\draw[->, color=blue] (da) to [out=-150, in=-30] node[anchor=north, xshift=-0.2cm] {$2$} (ba);
\draw[->, color=blue] (db) to [out=-150, in=-30] node[anchor=north, xshift=-0.2cm] {$2$} (bb);
\draw[->, color=blue] (dc) to [out=-150, in=-30] node[anchor=north, xshift=-0.2cm] {$2$} (bc);
\draw[->, color=blue] (dd) to [out=-150, in=-30] node[anchor=north, xshift=-0.2cm] {$2$} (bd);
\draw[->, color=blue] (de) to [out=-150, in=-30] node[anchor=north, xshift=-0.2cm] {$2$} (be);

\draw[->] (da) -- node[anchor=south] {$6$} (tb);
\draw[->] (db) -- node[anchor=south] {$6$} (tc);
\draw[->] (dc) -- node[anchor=south] {$6$} (td);
\draw[->] (dd) -- node[anchor=south] {$6$} (te);
\end{tikzpicture}
\caption{Section of the graph representing the dynamic programming scheme for the graph $G_0$ from Figure \ref{fig:nnEx}. Each edge is labeled with its associated resource consumption.} \label{fig:nnExDyn}
\end{center}
\end{figure}
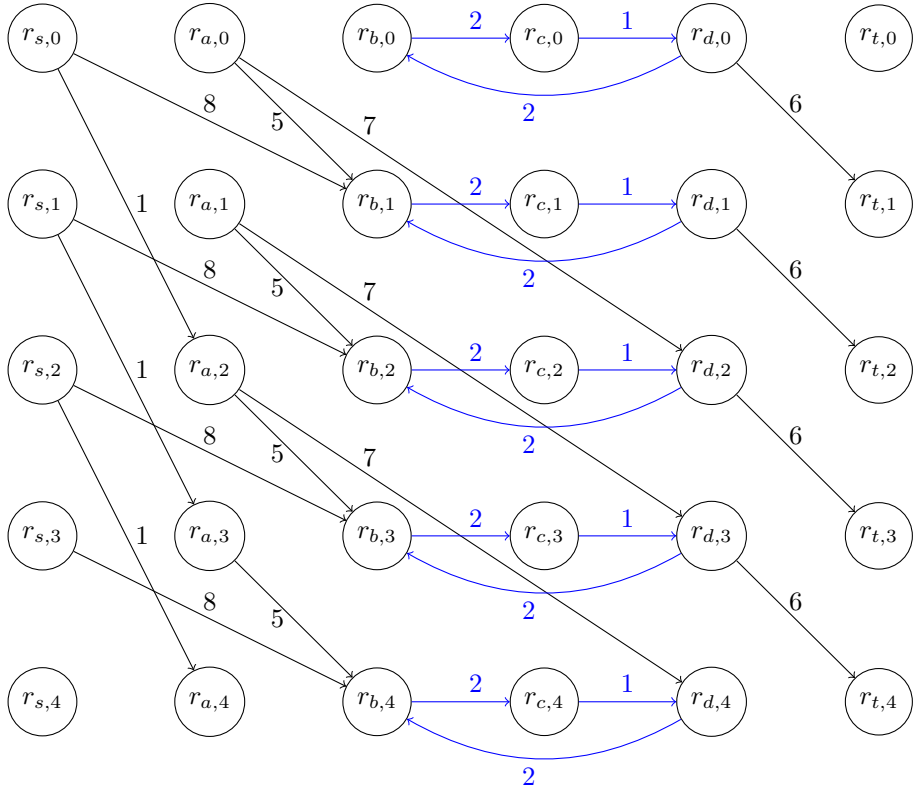

What can we do if there are edges $e$ with $c(e) = 0$? Figure \ref{fig:nnEx} shows an example of such a graph. Figure \ref{fig:nnExDyn} illustrates the dependencies between the values $r_{v, i}$. Because of the zero-cost edge cycle, there are cycles in this (infinite) graph, preventing us from computing these values independently.

Consider a path $p \in P_v$ minimizing $R_G(p)$ under the constraint $C_G(p) \leq i$. We have to distinguish two cases:
\begin{itemize}
\item If $C_G(p) > 0$, then $p$ can be split into paths $s \stackrel{p_1}{\longrightarrow} u \stackrel{p_2}{\longrightarrow} v$ where the last edge of $p_1$ has non-zero cost and all edges of $p_2$ have zero cost ($p_2$ might be empty). Note that we can assume $u \neq s$, since otherwise we could replace $p_1$ by an empty path without increasing cost or resource values. We can compute modified values
\begin{IEEEeqnarray*}{+rCl+x*}
r_{u, i}' & = & \min_{\substack{e = (w, u) \in E \\ c(e) > 0}} r_{w, i - c(e)} + r(e)
\end{IEEEeqnarray*}
that describe the minimum resource consumption among paths like $p_1$ from $s$ to $u$ with cost $\leq i$ and whose last edge has non-zero cost. Surely, $p_1$ must be the optimal path to $u$ and thus $R_G(p_1) = r_{u, i}'$. Finding the path from any such ``entry node'' $u$ minimizing the sum of $r_{u, i}$ and its own resource consumption must therefore yield the value of $r_{v, i}$. Consider the subgraph $G^{(i)}$ of $G$ containing only those edges $e$ with $c(e) = 0$. This is effectively a single-weighted graph, since only the resource consumption values may differ. This subgraph still contains all possible paths for $p_2$. To represent the paths $p_1$ for which we already know the possible lengths, we add a new vertex $v_s$ and edges $(v_s, u)$ with weights $r_{u, i}'$ for $u \neq s$.
\item If $C_G(p) = 0$, we cannot split $p$ into parts like above. But we know that $G^{(i)}$ contains all edges of $p$. Therefore, introducing an additional edge $(v_s, s)$ with weight zero in $G^{(i)}$ means that $p$ has the same resource consumption in $G$ as the path $v_s \longrightarrow s \stackrel{p}{\longrightarrow} v$ in $G^{(i)}$.
\end{itemize}

In $G^{(i)}$, each path from $v_s$ to $v$ corresponds to a path from $s$ to $v$ in $G$. Conversely, each minimum-resource path with cost $\leq i$ from $s$ to $v$ in $G$ corresponds to a path from $v_s$ to $v$ in $G^{(i)}$. Thus, the desired values $r_{v, i}$ are the lengths of the shortest paths $v_s \longrightarrow u \stackrel{p_2}{\longrightarrow} v$ from $v_s$ to $v$ in our modified subgraph $G^{(i)}$. This means we can apply any suitable single source shortest path (SSSP) algorithm to compute these values. 

Figure \ref{fig:nnExDynFilled} shows the graph from Figure \ref{fig:nnExDyn} before the values $r_{v, 4}$ are computed. The edges in red are relevant for computing the values $r_{v, 4}'$. These values are used in the modified graph $G_0^{(4)}$ shown in Figure \ref{fig:nnExModified}. 

\newcommand{\compNodeColor}{green!0!black}
\newcommand{\otherNodeColor}{orange!80!black}
\newcommand{\otherEdgeStyle}{dashed}
\newcommand{\minSize}{0.9cm}
\newcommand{\zeroEdgeColor}{blue}
\newcommand{\zeroEdgeStyle}{solid}

\begin{figure}[hbt]
\begin{center}
\begin{tikzpicture}[node distance = 2.2cm]
\node[minimum size = \minSize, color=\compNodeColor, circle, draw] (sa) {$0$};
\node[minimum size = \minSize, color=\compNodeColor, circle, draw, right of = sa] (aa) {$\infty$};
\node[minimum size = \minSize, color=\compNodeColor, circle, draw, right of = aa] (ba) {$\infty$};
\node[minimum size = \minSize, color=\compNodeColor, circle, draw, right of = ba] (ca) {$\infty$};
\node[minimum size = \minSize, color=\compNodeColor, circle, draw, right of = ca] (da) {$\infty$};
\node[minimum size = \minSize, color=\compNodeColor, circle, draw, right of = da] (ta) {$\infty$};
\node[minimum size = \minSize, color=\compNodeColor, circle, draw, below of = sa] (sb) {$0$};
\node[minimum size = \minSize, color=\compNodeColor, circle, draw, right of = sb] (ab) {$\infty$};
\node[minimum size = \minSize, color=\compNodeColor, circle, draw, right of = ab] (bb) {$8$};
\node[minimum size = \minSize, color=\compNodeColor, circle, draw, right of = bb] (cb) {$10$};
\node[minimum size = \minSize, color=\compNodeColor, circle, draw, right of = cb] (db) {$11$};
\node[minimum size = \minSize, color=\compNodeColor, circle, draw, right of = db] (tb) {$\infty$};
\node[minimum size = \minSize, color=\compNodeColor, circle, draw, below of = sb] (sc) {$0$};
\node[minimum size = \minSize, color=\compNodeColor, circle, draw, right of = sc] (ac) {$1$};
\node[minimum size = \minSize, color=\compNodeColor, circle, draw, right of = ac] (bc) {$8$};
\node[minimum size = \minSize, color=\compNodeColor, circle, draw, right of = bc] (cc) {$10$};
\node[minimum size = \minSize, color=\compNodeColor, circle, draw, right of = cc] (dc) {$11$};
\node[minimum size = \minSize, color=\compNodeColor, circle, draw, right of = dc] (tc) {$17$};
\node[minimum size = \minSize, color=\compNodeColor, circle, draw, below of = sc] (sd) {$0$};
\node[minimum size = \minSize, color=\compNodeColor, circle, draw, right of = sd] (ad) {$1$};
\node[minimum size = \minSize, color=\compNodeColor, circle, draw, right of = ad] (bd) {$6$};
\node[minimum size = \minSize, color=\compNodeColor, circle, draw, right of = bd] (cd) {$8$};
\node[minimum size = \minSize, color=\compNodeColor, circle, draw, right of = cd] (dd) {$9$};
\node[minimum size = \minSize, color=\compNodeColor, circle, draw, right of = dd] (td) {$17$};
\node[minimum size = \minSize, color=\otherNodeColor, circle, draw, below of = sd] (se) {$r_{s, 4}$};
\node[minimum size = \minSize, color=\otherNodeColor, circle, draw, right of = se] (ae) {$r_{a, 4}$};
\node[minimum size = \minSize, color=\otherNodeColor, circle, draw, right of = ae] (be) {$r_{b, 4}$};
\node[minimum size = \minSize, color=\otherNodeColor, circle, draw, right of = be] (ce) {$r_{c, 4}$};
\node[minimum size = \minSize, color=\otherNodeColor, circle, draw, right of = ce] (de) {$r_{d, 4}$};
\node[minimum size = \minSize, color=\otherNodeColor, circle, draw, right of = de] (te) {$r_{t, 4}$};

\draw[->, color = \compNodeColor] (sa) -- node[anchor=west] {$1$} (ac);
\draw[->, color = \compNodeColor] (sb) -- node[anchor=west] {$1$} (ad);
\draw[->, color = \otherNodeColor, \otherEdgeStyle] (sc) -- node[anchor=west] {$1$} (ae);

\draw[->, color = \compNodeColor] (sa) -- node[anchor=south] {$8$} (bb);
\draw[->, color = \compNodeColor] (sb) -- node[anchor=south] {$8$} (bc);
\draw[->, color = \compNodeColor] (sc) -- node[anchor=south] {$8$} (bd);
\draw[->, color = \otherNodeColor, \otherEdgeStyle] (sd) -- node[anchor=south] {$8$} (be);

\draw[->, color = \compNodeColor] (aa) -- node[anchor=east] {$5$} (bb);
\draw[->, color = \compNodeColor] (ab) -- node[anchor=east] {$5$} (bc);
\draw[->, color = \compNodeColor] (ac) -- node[anchor=east] {$5$} (bd);
\draw[->, color = \otherNodeColor, \otherEdgeStyle] (ad) -- node[anchor=east] {$5$} (be);

\draw[->, color = \compNodeColor] (aa) -- node[anchor=south, xshift=-1.2cm, yshift=0.8cm] {$7$} (dc);
\draw[->, color = \compNodeColor] (ab) -- node[anchor=south, xshift=-1.2cm, yshift=0.8cm] {$7$} (dd);
\draw[->, color = \otherNodeColor, \otherEdgeStyle] (ac) -- node[anchor=south, xshift=-1.2cm, yshift=0.8cm] {$7$} (de);

\draw[->, color=\compNodeColor] (ba) -- node[anchor=south, xshift=0.2cm] {$2$} (ca);
\draw[->, color=\compNodeColor] (bb) -- node[anchor=south, xshift=0.2cm] {$2$} (cb);
\draw[->, color=\compNodeColor] (bc) -- node[anchor=south, xshift=0.2cm] {$2$} (cc);
\draw[->, color=\compNodeColor] (bd) -- node[anchor=south, xshift=0.2cm] {$2$} (cd);
\draw[->, color=\zeroEdgeColor, \zeroEdgeStyle] (be) -- node[anchor=south, xshift=0.2cm] {$2$} (ce);

\draw[->, color=\compNodeColor] (ca) -- node[anchor=south] {$1$} (da);
\draw[->, color=\compNodeColor] (cb) -- node[anchor=south] {$1$} (db);
\draw[->, color=\compNodeColor] (cc) -- node[anchor=south] {$1$} (dc);
\draw[->, color=\compNodeColor] (cd) -- node[anchor=south] {$1$} (dd);
\draw[->, color=\zeroEdgeColor, \zeroEdgeStyle] (ce) -- node[anchor=south] {$1$} (de);

\draw[->, color=\compNodeColor] (da) to [out=-150, in=-30] node[anchor=north, xshift=-0.2cm] {$2$} (ba);
\draw[->, color=\compNodeColor] (db) to [out=-150, in=-30] node[anchor=north, xshift=-0.2cm] {$2$} (bb);
\draw[->, color=\compNodeColor] (dc) to [out=-150, in=-30] node[anchor=north, xshift=-0.2cm] {$2$} (bc);
\draw[->, color=\compNodeColor] (dd) to [out=-150, in=-30] node[anchor=north, xshift=-0.2cm] {$2$} (bd);
\draw[->, color=\zeroEdgeColor, \zeroEdgeStyle] (de) to [out=-150, in=-30] node[anchor=north, xshift=-0.2cm] {$2$} (be);

\draw[->, color = \compNodeColor] (da) -- node[anchor=south] {$6$} (tb);
\draw[->, color = \compNodeColor] (db) -- node[anchor=south] {$6$} (tc);
\draw[->, color = \compNodeColor] (dc) -- node[anchor=south] {$6$} (td);
\draw[->, color = \otherNodeColor, \otherEdgeStyle] (dd) -- node[anchor=south] {$6$} (te);
\end{tikzpicture}
\caption{Partially computed values in the infinite graph corresponding to $G_0$. The edges that are relevant for building $G_0^{(4)}$ and thus for computing the next row entries are dashed and highlighted in orange. Each edge is labeled with its associated resource consumption.} \label{fig:nnExDynFilled}
\end{center}
\end{figure}

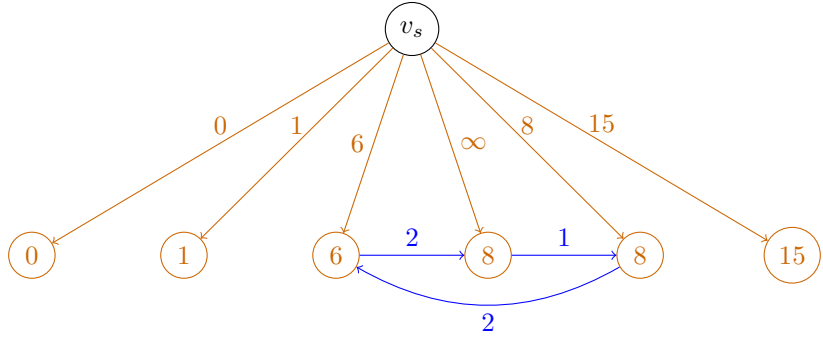
\begin{figure}[hbt]
\begin{center}
\begin{tikzpicture}[node distance = 2cm]
\node[circle, color=\otherNodeColor, draw] (s) {$0$};
\node[circle, color=\otherNodeColor, draw, right of = s] (a) {$1$};
\node[circle, color=\otherNodeColor, draw, right of = a] (b) {$6$};
\node[circle, color=\otherNodeColor, draw, right of = b] (c) {$8$};
\node[circle, color=\otherNodeColor, draw, right of = c] (d) {$8$};
\node[circle, color=\otherNodeColor, draw, right of = d] (t) {$15$};
\node[circle, draw, above of = b, node distance = 3cm, xshift=1cm] (vs) {$v_s$};

\draw[->, color=\zeroEdgeColor] (b) -- node[anchor=south] {$2$} (c);
\draw[->, color=\zeroEdgeColor] (c) -- node[anchor=south] {$1$} (d);
\draw[->, color=\zeroEdgeColor] (d) to [out=-150, in=-30] node[anchor=north] {$2$} (b);
\draw[->, color=\otherNodeColor] (vs) -- node[anchor=south] {$0$} (s);
\draw[->, color=\otherNodeColor] (vs) -- node[anchor=south] {$1$} (a);
\draw[->, color=\otherNodeColor] (vs) -- node[anchor=east] {$6$} (b);
\draw[->, color=\otherNodeColor] (vs) -- node[anchor=west] {$\infty$} (c);
\draw[->, color=\otherNodeColor] (vs) -- node[anchor=south] {$8$} (d);
\draw[->, color=\otherNodeColor] (vs) -- node[anchor=south] {$15$} (t);
\end{tikzpicture}
\caption{The graph $G_0^{(4)}$ and its shortest path weight entries corresponding to the values $r_{4, v}$. Each edge is labeled with its associated cost.} \label{fig:nnExModified}
\end{center}
\end{figure}

Unless we have sufficiently many edges\footnote{i.e. $m \in \compGeq{n \log n}$} or a special instance like a directed acyclic graph where faster algorithms exist, we now have a slightly increased time complexity of $\compLeq{(m + n \log n)(1 + \copt(G))}$ using Dijkstra's algorithm with a Fibonacci heap. Additionally, we could decide to interrupt the algorithm if we did not find $\copt(G)$ after having computed the $b$-th row for some value of $b$. To abstract from the reduced runtime in special cases, we propose the following definition:

\begin{definition} \label{def:rowcost}
We denote by $\rowtime{G}(n, m)$ a function which represents the time complexity needed for computing one row of the dynamic programming scheme described above, using $G$ with any modified cost function $c: E \to \natZero$. For example we might set $f_G(n, m) \equalDef m$ for acyclic graphs $G$ and $f_G(n, m) \equalDef m + n \log n$ for general graphs. Further options are presented at the end of this paper. 
\end{definition}

The results of this section can now be summarized in the following lemma:
\begin{lemma} \label{lemma:optimal:complexity}
If $c(e) \in \natZero$ for all $e \in E$, there exists an algorithm $\textsc{ExactRSP}(G, s, t, R, b)$ which returns
\begin{itemize}
\item $\copt(G)$ (and optionally $\popt(G)$), if $\copt(G) \leq b$ and
\item $\infty$, if $\copt(G) \geq b+1$
\end{itemize}
with a time complexity of
\begin{itemize}
\item $\compLeq{m(1 + \min\{b, \copt(G)\})}$, if $c(e) \neq 0$ for all $e \in E$, or
\item $\compLeq{\rowtime{G}(n, m) \cdot (1 + \min\{b, \copt(G)\})}$ otherwise.
\end{itemize}
\end{lemma}

\section{Approximation Schemes}

Since finding an optimal solution to RSP is an NP-hard problem \cite{garey1979}, we want to find a solution that approximates the optimum in polynomial time. Concretely, for some given $\varepsilon > 0$, we want to find a $(1 + \varepsilon)$-approximation, that is, a path $p \in P_t$ satisfying $R_G(p) \leq R$ and $C_G(p) \leq (1 + \varepsilon) \copt(G)$.

\subsection{Scaling}

For any scaling factor $S > 0$, we can build new graphs $\floorGraph{S} \equalDef (V, E, \floorCost{S}, r)$ and $\ceilGraph{S} \equalDef (V, E, \ceilCost{S}, r)$ by using scaled and rounded costs
\begin{IEEEeqnarray*}{+rCl+x*}
\floorCost{S}(e) & \equalDef & \left \lfloor \frac{c(e)}{S} \right \rfloor \\
\ceilCost{S}(e) & \equalDef & \left \lfloor \frac{c(e)}{S} \right \rfloor + 1~.
\end{IEEEeqnarray*}

\begin{lemma} \label{lemma:scaledCost}
For any scaling factor $S > 0$, we have
\begin{IEEEeqnarray*}{+rCl+x*}
\frac{\copt(G)}{S} - n \leq \copt(\floorGraph{S}) \leq \frac{\copt(G)}{S} \leq \copt(\ceilGraph{S}) \leq \frac{\copt(G)}{S} + n~.
\end{IEEEeqnarray*}

\begin{proof}
The first inequality follows from 
\begin{IEEEeqnarray*}{+rCl+x*}
\copt(\floorGraph{S}) & = & \sum_{e \in \popt(\floorGraph{S})} \left \lfloor \frac{c(e)}{S} \right \rfloor \geq \sum_{e \in \popt(\floorGraph{S})} \left( \frac{c(e)}{S} - 1 \right) \\
& \geq & \frac{1}{S}\left(\sum_{e \in \popt(\floorGraph{S})} c(e) \right) - n \geq \frac{1}{S}\left(\sum_{e \in \popt(G)} c(e) \right) - n \geq \frac{\copt(G)}{S} - n~.
\end{IEEEeqnarray*}
Furthermore, we obtain
\begin{IEEEeqnarray*}{+rCl+x*}
\copt(\floorGraph{S}) & = & \sum_{e \in \popt(\floorGraph{S})} \left \lfloor \frac{c(e)}{S} \right \rfloor \leq \sum_{e \in \popt(G)} \left \lfloor \frac{c(e)}{S} \right \rfloor \leq \sum_{e \in \popt(G)} \frac{c(e)}{S} = \frac{\copt(G)}{S}~.
\end{IEEEeqnarray*}
The remaining inequalities can be shown similarly.
\end{proof}
\end{lemma}

\subsection{Previous Results}
The results following in this subsection were published in \cite{lorenz2001}. Because they are essential to our algorithm, we restate them here in a form adapted to our improved algorithms.

\begin{lemma} \label{lemma:scaling:approximation}
For any scaling factor $S > 0$, we have
\begin{IEEEeqnarray*}{+rCl+x*}
C_G(\popt(\ceilGraph{S})) \leq \copt(G) + Sn
\end{IEEEeqnarray*}

\begin{proof}
By Lemma \ref{lemma:scaledCost} and the definition of $C_G$ and $C_{\ceilGraph{S}}$, we have
\begin{IEEEeqnarray*}{+rCl+x*}
C_G(\popt(\ceilGraph{S})) & \leq & S \cdot C_{\ceilGraph{S}}(\popt(\ceilGraph{S})) = S \cdot \copt(\ceilGraph{S}) \\
& \leq & S \cdot \left( \frac{\copt(G)}{S} + n\right) = \copt(G) + Sn~. & \qedhere
\end{IEEEeqnarray*}
\end{proof}
\end{lemma}

\begin{lemma} \label{lemma:scaling:runtime}
Given a value $L$ satisfying $\frac{\copt(G)}{d} \leq L \leq \copt(G)$ for a constant $d \geq 1$ independent of $L$, we can compute a $(1 + \varepsilon)$-approximation in time $\compLeq{nm/\varepsilon}$.

\begin{proof}
We choose $S \equalDef \varepsilon \cdot L / n$. By Lemma \ref{lemma:scaling:approximation}, $\popt(\ceilGraph{S})$ is a $(1 + \varepsilon)$-approximation. By Lemma \ref{lemma:optimal:complexity} and Lemma \ref{lemma:scaledCost}, we can find $\popt(\ceilGraph{S})$ in time
\begin{IEEEeqnarray*}{+rCl+x*}
\compLeq{m \left(\frac{\copt(G)}{S} + n + 1 \right)} & \subseteq & \compLeq{m\left( \frac{dn}{\varepsilon} + n + 1 \right)} = \compLeq{nm/\varepsilon }. & \qedhere
\end{IEEEeqnarray*}
\end{proof}
\end{lemma}

Finding a value $L$ as used in Lemma \ref{lemma:scaling:runtime} is not trivial. In the following, we want to derive a method to find a constant approximation (i.e. lower and upper bounds $L, U$ for $\copt(G)$ with $U/L \in \compLeq{1}$.

\begin{lemma} \label{lemma:approximation:linear}
Lower and upper bounds $L, U$ for $\copt(G)$ satisfying $U = nL$ can be found with a time complexity of $\compLeq{(m + n \log n) \log m}$.
\begin{proof}
Let $\{e_1, \hdots, e_m\}$ be the edges of $G$ sorted ascending by $c(e)$. Using Dijkstra's algorithm, we can check if there is a path $p$ from $s$ to $t$ satisfying $R_G(p) \leq R$ which only uses edges from $\{e_1, \hdots, e_j\}$. Let $j^*$ be the smallest value of $j$ such that there is such a path. We can find $j^*$ using binary search on $j$ in time $\compLeq{(m + n \log n) \log m}$. Because there is no such path for $j^* - 1$, each valid path must have at least one edge with cost $c(e_{j^*})$. Thus, $L \equalDef c(e_{j^*})$ is a lower bound for $\copt(G)$. Since there is a path of length at most $n$ using only edges from $\{e_1, \hdots, e_{j^*}\}$, $U \equalDef n \cdot c(e_{j^*})$ is an upper bound for $\copt(G)$. Thus, we can find upper and lower bounds with $U = nL$ in the specified time.
\end{proof}
\end{lemma}

As a side effect, the algorithm from Lemma \ref{lemma:approximation:linear} allows us to efficiently detect the trivial case where $\copt(G) = 0$. Thus, we will assume $\copt(G) > 0$ in the following. This allows us to divide by $\copt(G)$ or by appropriate bounds for it.

\begin{definition} \label{def:scalingFactors}
Let $L, U$ be the bounds provided by Lemma \ref{lemma:approximation:linear}. Then we define the scaling factors $S_i \equalDef 2^{-i} U/(2n)$ for $i \in \natZero$. The names $L, U, S_i$ are used in the remainder of the document.
\end{definition}

\begin{lemma} \label{lemma:numScalingFactors}
Let $L, U$ be lower and upper bounds for $\copt(G)$ and let $k \equalDef \lceil \log(U/L) \rceil$. Then we have $\copt(G)/S_0 \leq 2n$ and $\copt(G)/S_k \geq 2n$.

\begin{proof}
We have
\begin{IEEEeqnarray*}{+rCl+x*}
\frac{\copt(G)}{S_0} & = & \frac{\copt(G)}{U} \cdot 2n \leq 1 \cdot 2n \\
\frac{\copt(G)}{S_k} & \geq & \frac{\copt(G)}{\frac{L}{U} \cdot \frac{U}{2n}} = 2n \cdot \frac{\copt(G)}{L} \geq 2n~. & \qedhere
\end{IEEEeqnarray*}
\end{proof}
\end{lemma}

\begin{lemma} \label{lemma:binarySearch}
Given lower and upper bounds $L, U$ for $\copt(G)$, we can find new bounds $L', U'$ satisfying $U'/L' \in \compLeq{1}$ in time $\compLeq{nm \log \log(U/L)}$.

\begin{proof}
Using $\textsc{ExactRSP}(\ceilGraph{S_i}, s, t, R, 5n)$, we can test if $\copt(\ceilGraph{S_i}) < 2n$, $\copt(\ceilGraph{S_i}) > 5n$ or $\copt(\ceilGraph{S_i}) \in [2n, 5n]$ in time $\compLeq{mn}$ (Lemma \ref{lemma:optimal:complexity}). By Lemma \ref{lemma:scaledCost}, the latter is definitely the case if $\copt(G)/S_i \in [2n, 4n]$. This in turn is, by Lemma \ref{lemma:numScalingFactors}, surely the case for at least one $i \in \{0, \hdots, \lceil \log(U/L) \rceil\}$. Using binary search on $i$, we can therefore find a value of $i$ satisfying $\copt(\ceilGraph{S_i}) \in [2n, 5n]$ in time $\compLeq{nm \log \log(U/L)}$. Given this particular value of $i$, we then know from Lemma \ref{lemma:scaledCost} that
\begin{IEEEeqnarray*}{+rCl+x*}
L' & \equalDef & S_i \left(\copt(\ceilGraph{S_i}) - n\right) \leq S_i \cdot \frac{\copt(G)}{S_i} = \copt(G) \\
U' & \equalDef & S_i \cdot \copt(\ceilGraph{S_i}) \geq S_i \cdot \frac{\copt(G)}{S_i} = \copt(G)
\end{IEEEeqnarray*}
are lower and upper bounds for $\copt(G)$ with
\begin{IEEEeqnarray*}{+rCl+x*}
\frac{U'}{L'} & = & \frac{\copt(\ceilGraph{S_i})}{\copt(\ceilGraph{S_i}) - n} \leq \frac{2n}{2n - n} = 2 \in \compLeq{1}~. & \qedhere
\end{IEEEeqnarray*}
\end{proof}
\end{lemma}

\begin{theorem}[Lorenz, Raz]
A $(1 + \varepsilon)$-ap\-pro\-xi\-ma\-tion to the optimal solution can be found with a time complexity of $\compLeq{nm\left(\frac{1}{\varepsilon} + \log \log n\right)}$.

\begin{proof}
This is a direct consequence of Lemmas \ref{lemma:scaling:runtime}, \ref{lemma:approximation:linear} and \ref{lemma:binarySearch}.
\end{proof}
\end{theorem}

\subsection{Main Results} \label{subsection:mainResults}

In this section, we present our new algorithm using linear search to find bounds $U', L'$ with $U'/L' \in \compLeq{1}$ for $\copt(G)$. In particular, our goal is to find the smallest $i \in \natZero$ such that $\copt(\floorGraph{S_i}) > b$ for some value of $b \in \natOne$. The following lemma states why this is useful to obtain tighter bounds for $\copt(G)$. Subsequently, we present our algorithm and its analysis.

\begin{lemma} \label{lemma:search:quality}
Let $i^*$ be the lowest value for $i \geq 0$ such that $\copt(\floorGraph{S_i}) > b$ (we will prove its existence in Lemma \ref{lemma:maxIterations}). We define
\begin{IEEEeqnarray*}{+rCl+x*}
L' & \equalDef & b \cdot S_{i^*} \\
U' & \equalDef & \begin{cases}
U &, i^* = 0 \\
2S_{i^*}(b + n) &, i^* > 0~.
\end{cases}
\end{IEEEeqnarray*}
Then, $(L', U')$ are correct lower and upper bounds for $\copt(G)$ with $U'/L' \in \compLeq{1 + \frac{n}{b}}$.

\begin{proof}
Using Lemma \ref{lemma:scaledCost}, we obtain $L' = b \cdot S_{i^*} \leq \copt(\floorGraph{S_{i^*}}) \cdot S_{i^*} \leq \copt(G)$. By definition, $U$ is an upper bound for $\copt(G)$. If $i^* > 0$, we know that $\copt(\floorGraph{S_{i^* - 1}}) \leq b$. Furthermore, we know $S_{i^* - 1} = 2S_{i^*}$ by definition. Again by Lemma \ref{lemma:scaledCost}, we obtain 
\begin{IEEEeqnarray*}{+rCl+x*}
2S_{i^*}(b + n) & \geq & S_{i^* - 1}\left( \copt(\floorGraph{S_{i^* - 1}}) + n \right) \geq S_{i^* - 1} \cdot \frac{\copt(G)}{S_{i^* - 1}} = \copt(G)~.
\end{IEEEeqnarray*}
For the quotient $U'/L'$, we have
\begin{IEEEeqnarray*}{+rCl+x*}
\frac{U'}{L'} & = & \frac{U}{b \cdot \frac{U}{2n}} = \frac{2n}{b} \in \compLeq{1 + \frac{n}{b}}
\end{IEEEeqnarray*}
if $i^* = 0$ and
\begin{IEEEeqnarray*}{+rCl+x*}
\frac{U'}{L'} & = & \frac{2S_{i^*}(b + n)}{b \cdot S_{i^*}} = 2 + 2 \cdot \frac{n}{b} \in \compLeq{1 + \frac{n}{b}}
\end{IEEEeqnarray*}
if $i^* > 0$.
\end{proof}
\end{lemma}

\begin{lemma} \label{lemma:halving}
For any $i \geq 0$, $\copt(\floorGraph{S_{i+1}}) \geq 2 \cdot \copt(\floorGraph{S_i})$. By repeatedly applying this inequality, we can conclude that for any $i < i^*$,
\begin{IEEEeqnarray*}{+rCl+x*}
\copt(\floorGraph{S_i}) \leq 2^{-(i^* - 1 - i)} \copt(\floorGraph{S_{i^* - 1}}) \leq 2^{i + 1 - i^*} \cdot b~.
\end{IEEEeqnarray*}

\begin{proof}
For any $i \geq 0$ and any edge $e \in E$, we have
\begin{IEEEeqnarray*}{+rCl+x*}
\floorCost{S_{i+1}}(e) & = & \left \lfloor \frac{c(e)}{S_{i+1}} \right \rfloor = \left \lfloor 2 \cdot \frac{c(e)}{S_i} \right \rfloor \geq 2 \cdot \left \lfloor \frac{c(e)}{S_i} \right \rfloor = 2 \cdot \floorCost{S_i}(e)~.
\end{IEEEeqnarray*}
Thus, $\copt(\floorGraph{S_{i+1}}) \geq 2 \cdot C_{\floorGraph{S_i}} \left( \popt(\floorGraph{S_{i+1}}) \right) \geq 2 \cdot \copt(\floorGraph{S_i})$.
\end{proof}
\end{lemma}

Lemma \ref{lemma:halving} enables us to use linear search efficiently to find the value $i^*$ from Lemma \ref{lemma:search:quality}. The details are shown in Algorithm \ref{alg:linearSearch} that uses the algorithm \textsc{ExactRSP} from Lemma \ref{lemma:optimal:complexity}.

\begin{algorithm}[hbt]
\begin{algorithmic}[1]
\Function{LSBounds}{$G = (V, E, c, r), s \in V, t \in V, R \in \nonNegReals, b \in \natOne$}
	\State Use $L, U, S_i$ from Definition \ref{def:scalingFactors}
	\For{$i = 0, 1, 2, \hdots$}
		\State $S \assign S_i$
		\If{\Call{ExactRSP}{$\floorGraph{S}, s, t, R, b$} $= \infty$}
			\If{$i = 0$}
				\State \Return $(S b, U)$ \label{LSBounds:FirstCase}
			\Else
				\State \Return $(S b, 2S (b+n))$ \label{LSBounds:SecondCase}
			\EndIf
		\EndIf
	\EndFor 
\EndFunction
\end{algorithmic}
\caption{Algorithm using linear search to find bounds for $\copt(G)$} \label{alg:linearSearch}
\end{algorithm}

In order to analyze the runtime of the algorithm, we need another lemma.

\begin{lemma} \label{lemma:maxIterations}
If $b \leq n$, $i^* \leq \lceil \log n \rceil$ holds.

\begin{proof}
Assuming $b \leq n$, we obtain
\begin{IEEEeqnarray*}{+rCl+x*}
\copt(\floorGraph{S_{\lceil \log n \rceil}}) & \geq & \frac{\copt(G)}{S_{\lceil \log n \rceil}} - n & (Lemma \ref{lemma:scaledCost}) \\
& \geq & \frac{\copt(G)}{\frac{1}{n} \cdot \frac{U}{2n}} - n \\
& \geq & 2n \cdot \frac{\copt(G)}{\copt(G)} - n & (Lemma \ref{lemma:approximation:linear}) \\
& = & n \geq b~. & \qedhere
\end{IEEEeqnarray*}
\end{proof}
\end{lemma}

Now we can prove

\begin{theorem} \label{thm:improved}
For any $b \leq n$, executing the statement $(L', U') \equalDef \textsc{LSBounds}(G, s, t, R, b)$ provides lower and upper bounds satisfying $U'/L' \in \compLeq{1 + \frac{n}{b}}$ with a time complexity of $\compLeq{f_G(n, m) \cdot (1 + b + \log n)}$.\footnote{It turns out that the assumption $b \leq n$ is not necessary, but it makes Lemma \ref{lemma:maxIterations} easier.}

\begin{proof}
The estimation quality claim has already been proven in Lemma \ref{lemma:search:quality}. Obviously, the runtime of the function \textsc{ExactRSP} dominates the runtime of the algorithm (since the initial bounds $L, U$ from Lemma \ref{lemma:approximation:linear} can be found quite efficiently). From Lemma \ref{lemma:optimal:complexity}, we know that we can bound the runtime of $\textsc{ExactRSP}(\floorGraph{S}, s, t, R, b)$ by
\begin{IEEEeqnarray*}{+rCl+x*}
Af_G(n, m) \cdot (1 + \min\{b, \copt(\floorGraph{S})\}) + D
\end{IEEEeqnarray*}
for some constants $A, D \in \posReals$. For the cumulative runtime $T$ of all calls to \textsc{ExactRSP}, we can derive
\begin{IEEEeqnarray*}{+rCl+x*}
T & \in & A f_G(n, m) b \left( \left(\sum_{i=0}^{i^*-1} 2^{i + 1 - i^*} \right) + 1 \right) + \, \compLeq{f_G(n, m) \cdot (1 + i^*)} & (Lemma \ref{lemma:halving}) \\
& \subseteq & A f_G(n, m) b (2 + 1) + \compLeq{f_G(n, m) \cdot (1 + \log n)} & (Lemma \ref{lemma:maxIterations}) \\
& \subseteq & \compLeq{f_G(n, m) (1 + b + \log n)}~. & \qedhere
\end{IEEEeqnarray*}
\end{proof}
\end{theorem}

\begin{corollary} \label{cor:nonlogresult}
A $(1 + \varepsilon)$-approximation for RSP can be found in time $\compLeq{nm \left( \frac{1}{\varepsilon} + \frac{f_G(n, m)}{m} \right)}$.

\begin{proof}
By Theorem \ref{thm:improved}, executing the statement $(L', U') \equalDef \textsc{LSBounds}(G, s, t, R, b)$ with $b \equalDef n$ yields bounds satisfying
\begin{IEEEeqnarray*}{+rCl+x*}
\frac{U'}{L'} \in \compLeq{1 + n/b} = \compLeq{1}
\end{IEEEeqnarray*}
in time
\begin{IEEEeqnarray*}{+rCl+x*}
\compLeq{\rowtime{G}(n, m) \cdot (1 + b + \log n)} & = & \compLeq{n \cdot \rowtime{G}(n, m)}~.
\end{IEEEeqnarray*}
By Lemma \ref{lemma:scaling:runtime}, we can then find the desired $(1 + \varepsilon)$-approximation in $\compLeq{nm/\varepsilon}$.
\end{proof}
\end{corollary}

\begin{remark}
If $G$ is an undirected graph with positive integer resource values, we can use an algorithm by Thorup (working on the RAM model) to solve the SSSP problem in $\compLeq{m}$, assuming constant-time multiplication \cite{thorup1999}. This yields an overall runtime for the FPTAS of $\compLeq{nm/\varepsilon}$, just as for acyclic graphs where Ergun et al. have already proposed a slightly different algorithm with the same time complexity \cite{ergun2002}. If we use Dijkstra's algorithm with the priority queue proposed by van Emde Boas et al. \cite{van_emde_boas1976}, which is possible if $r(e) \in \natZero$ for all $e \in E$, we can use $f_G(n, m) \equalDef m + n \log \log R$ in Corollary \ref{cor:nonlogresult}.
\end{remark}

The following corollary provides an alternative algorithm to Xue et al. \cite{xue2008}.

\begin{corollary} \label{cor:logresult}
A $(1 + \varepsilon)$-approximation for RSP can be found in time $\compLeq{nm \left(\frac{1}{\varepsilon} + \log \log \log n \right)}$.

\begin{proof}
By Theorem \ref{thm:improved}, executing the statement $(L', U') \equalDef \textsc{LSBounds}(G, s, t, R, b)$ with
\begin{IEEEeqnarray*}{+rCl+x*}
b \coloneqq \left \lfloor \frac{n}{\log n} \right \rfloor
\end{IEEEeqnarray*}
yields bounds satisfying
\begin{IEEEeqnarray*}{+rCl+x*}
\frac{U'}{L'} \in \compLeq{1 + n/b} = \compLeq{1 + \log n} = \compLeq{\log n}
\end{IEEEeqnarray*}
in time
\begin{IEEEeqnarray*}{+rCl+x*}
&& \compLeq{\rowtime{G}(n, m) \cdot (1 + b + \log n)} \\
& \subseteq & \compLeq{(m + n \log n) \cdot \left(1 + \frac{n}{\log n} + \log n\right)} \\
& = & \compLeq{n \left(\frac{m}{\log n} + n\right)}~,
\end{IEEEeqnarray*}
and thus because of $m \geq n-1$ also in time $\compLeq{nm}$. By Lemma \ref{lemma:binarySearch}, we then can find bounds $L'', U''$ with a constant quotient in time $\compLeq{nm \log \log \log n}$. The actual path computation can then happen in time $\compLeq{nm/\varepsilon}$ according to Lemma \ref{lemma:scaling:runtime}.
\end{proof}
\end{corollary}

\paragraph*{Planar Graphs}
Henzinger et al. \cite{henzinger1997} showed that the SSSP problem for planar graphs can be solved in linear time. In the special cases we just presented, we implicitly used the fact that all mentioned properties like acyclicity easily translate from $G$ to $G^{(i)}$. Unfortunately, this is not the case for planarity. More specifically, $G^{(i)}$ is planar if and only if the zero-cost-edge-subgraph of $G$ is outerplanar. Hence, we have to verify some of the more general conditions in \cite{henzinger1997} to show that we can speed up our algorithm for all planar graphs $G$.

\begin{definition}[\cite{lipton1979}, \cite{henzinger1997}] \label{def:separable}
Let $\calC$ be a subgraph-closed class of graphs and $f: \natZero \to \bbR$ a function\footnote{Of course, $f$-separability is only nontrivial for sublinear functions $f$.}. $\calC$ is said to be \emph{$f$-separable} if constants $\alpha < 1, \beta > 0$ exist such that for every $n$-vertex graph $G = (V, E) \in \calC$, $V$ is the disjoint union of three subsets $A, B, C$ such that $|A|, |B| \leq \alpha n$, $|C| \leq \beta f(n)$ and no edge connects vertices from $A$ and $B$.
$\calC$ is said to fulfil an $f$-separator theorem iff $\calC$ is $f$-separable.
\end{definition}

\begin{lemma} \label{lemma:almostPlanarSeparable}
Let $\calR$ be the class of all graphs that can be made planar by removing one node\footnote{$\calR$ thus contains all planar graphs.}, including the empty graph. Then, $\calR$ is subgraph-closed, minor-closed and $\sqrt{n}$-separable.

\begin{proof}
The claim is trivial for the empty graph. For all other $G = (V, E) \in \calR$, let $p(G)$ denote a node such that after removing $p(G)$ from $G$, the remaining graph $\widehat{G}$ is planar. Let $G' = (V', E')$ be a subgraph of $G$, i.e. $V' \subseteq V$ and $E' \subseteq E \cap (V' \times V')$. If $p(G) \not \in V'$, $G'$ is a subgraph of the planar graph $\widehat{G}$ and thus already planar. When removing any node, it will stay planar. If $p(G) \in V'$, then we obtain a planar graph by removing $p(G)$ from $G'$ as we just explained.

A minor of $G$ can be obtained from a subgraph of $G$ by edge contractions. Since we have already shown that any subgraph of $G$ is in $\calR$, we only have to show that $\calR$ is closed under edge contractions. Let $G' = (V', E')$ be the graph that results from contracting the edge $e \in E$ in $G$. If $e = (v, p(G))$ or $e = (p(G), v)$, removing the newly formed node from $G'$ yields the subgraph of $\widehat{G}$ obtained by removing $v$. On the other hand, if $e$ is not incident to $p(G)$, removing $p(G)$ from $G'$ yields a minor of $\widehat{G}$. Because planar graphs are minor-closed, it follows that $G' \in \calR$. Therefore, $\calR$ is minor-closed.

Lipton and Tarjan \cite{lipton1979} showed that planar graphs are $\sqrt{n}$-separable, so they satisfy Definition \ref{def:separable} for some constants $\alpha', \beta'$. Let $\alpha \equalDef \alpha', \beta \equalDef \beta' + 1$. Consider an arbitrary graph $G = (V, E) \in \calR$. Since $\widehat{G}$ is planar, we can find a disjoint partition $V = A' \cup B' \cup C' \cup \{p(G)\}$ such that there is no edge joining vertices from $A$ and $B$ and we have $|A|, |B| \leq \alpha' (|V| - 1)$ and $|C| \leq \beta' \sqrt{|V| - 1}$. Using $A \equalDef A', B \equalDef B', C \equalDef C' \cup \{p(G)\}$ and the new constants $\alpha$ and $\beta$, we can easily see that the conditions from Definition \ref{def:separable} are met. Thus, $\calR$ is $\sqrt{n}$-separable.
\end{proof}
\end{lemma}

\begin{corollary} \label{cor:planarGraphRuntime}
If $G$ is planar, a $(1 + \varepsilon)$-approximation for RSP can be found in time $\compLeq{nm/\varepsilon}$.

\begin{proof}
By Lemma \ref{lemma:almostPlanarSeparable}, $\calR$ satisfies the conditions from Theorem 3.23 and section 4 in \cite{henzinger1997}, which means that the SSSP problem for graphs in $\calR$ can be solved in linear time. Since removing $v_s$ from any graph $G^{(i)}$ yields a (planar) subgraph of $G$, the SSSP problem on the graphs $G^{(i)}$ can be solved in linear time. Since this means that we can set $\rowtime{G}(n, m) = m \in \Theta(n)$, the claim follows from Corollary \ref{cor:nonlogresult}.
\end{proof}
\end{corollary}

\section{Conclusions}

In this paper, we have presented an algorithm for computing a $(1 + \varepsilon)$-approximation to the RSP problem in $\compLeq{nm\left(1/\varepsilon + \log \log \log n\right)}$. While it can be considered simpler than the previous algorithm achieving this runtime for general graphs, we also show how to adapt our algorithm to special graph classes, e.g. planar graphs, achieving the best currently known running time of $\compLeq{nm/\varepsilon}$.

It remains an open problem whether a runtime of $\compLeq{nm/\varepsilon}$ can be achieved for general graphs.



\nocite{*}
\bibliography{references}




\end{document}